\documentclass{ifacconf}

\usepackage{graphicx}      
\usepackage{natbib}        

\usepackage{graphicx}

\usepackage{amsmath}
\usepackage{amssymb}
\usepackage{amsfonts}
\usepackage{times}
\usepackage{mathptmx} 
\usepackage{datetime}

\newcommand{\sinc}{\mathrm{sinc}}
\newcommand{\tanc}{\mathrm{tanc}}
\newcommand{\tanhc}{\mathrm{tanhc}}

\DeclareMathAlphabet{\bit}{OML}{cmm}{b}{it}

\usepackage{datetime}
\usepackage{times} 
\usepackage{framed} 
\usepackage{graphicx} 
\usepackage{amsmath} 
\usepackage{amssymb}  

\def\fH{\mathfrak{H}}

\def\<{\leqslant}           
\def\>{\geqslant}           

\def\d{\partial}

\def\Re{\mathrm{Re}}   
\def\Im{\mathrm{Im}}   

\def\cH{\mathcal{H}}   
\def\mR{\mathbb{R}}    
\def\mC{\mathbb{C}}    

\def\Tr{\mathrm{Tr}}       
\def\rT{\mathrm{T}}        

\def\bE{\mathbf{E}}    

\def\[[[{[\![\![}
\def\]]]{]\!]\!]}

\def\bra{{\langle}}
\def\ket{{\rangle}}

\def\re{\mathrm{e}}        
\def\rd{\mathrm{d}}        

\def\cL{\mathcal{L}}

\def\bJ{\mathbf{J}}

\def\x{\times}
\def\ox{\otimes}

\def\fF{{\mathfrak F}}

\def\fP{\mathfrak{P}}

\def\cZ{\mathcal{Z}}
\def\fZ{\mathfrak{Z}}

\def\sH{\mathsf{H}}

\def\cX{\mathcal{X}}

\def\cK{\mathcal{K}}

\def\cV{\mathcal{V}}
\def\cC{\mathcal{C}}

\def\sP{\mathsf{P}}
\def\sQ{\mathsf{Q}}

\def\cI{\mathcal{I}}
\def\cP{{\mathcal P}}

\def\cA{\mathcal{A}}
\def\cB{\mathcal{B}}

\def\eps{\epsilon}
\def\Ups{\Upsilon}

\usepackage{datetime}

\begin{document}
\begin{frontmatter}
\title{\large Measurement-based Feedback Control of Linear Quantum Stochastic Systems with Quadratic-Exponential 
Criteria\thanksref{footnoteinfo}}%
\thanks[footnoteinfo]{This work is supported by the Air Force Office of Scientific Research (AFOSR) under agreement number FA2386-16-1-4065 and the Australian Research Council under grant DP180101805.}

\author[IGV]{Igor G. Vladimirov$^*$, \qquad Matthew R. James$^*$, \qquad Ian R. Petersen}

\address[IGV]{Research School of Electrical, Energy and Materials Engineering, College of Engineering and Computer Science,
Australian National University, Canberra, Acton, ACT 2601,
Australia (e-mail: igor.g.vladimirov@gmail.com, matthew.james@anu.edu.au, i.r.petersen@gmail.com).}

\begin{abstract}                
This paper is concerned with a risk-sensitive  optimal control  problem for a feedback connection of a quantum plant with a measurement-based  classical controller. The plant is a multimode open quantum harmonic oscillator driven by a multichannel  quantum Wiener process, and the controller is a linear time invariant system governed by a stochastic differential equation. The control objective is to stabilize the closed-loop system and minimize the infinite-horizon  asymptotic growth rate of  a quadratic-exponential functional (QEF) which penalizes the plant variables and the controller output. We combine a frequency-domain representation of the QEF growth rate, obtained recently,  with variational techniques and establish first-order necessary conditions of optimality for the state-space matrices  of the controller.
\end{abstract}


\end{frontmatter}

%

\section{Introduction}

The dynamics of open quantum systems  \cite{M_1998,GZ_2004,S_1994} are substantially influenced by their interaction with environment, which can include other quantum systems, electromagnetic radiation and classical (for example, mechanical) systems. This interaction is accompanied by energy exchange (and dissipation) and modifies the statistical properties of dynamic variables, which are described in quantum probabilistic terms \cite{H_2001,H_2018,M_1995} and are affected by noncommutativity. Such properties of the system variables are often quantified by their moments whose
moderate values are relevant for maintaining the system  in a regime  where its approximate (for example, linearized) model remains satisfactory.

In particular,  second-order moments of system variables provide mean square optimality criteria for linear quadratic Gaussian (LQG) quantum control and filtering problems \cite{B_1983,BVJ_2007,EB_2005,MP_2009,MJ_2012,NJP_2009,WM_2010,ZJ_2012}. These settings are concerned with linear quantum stochastic systems, or open quantum harmonic oscillators (OQHOs), which are modelled in the framework of the Hudson-Partha\-sa\-rathy calculus \cite{HP_1984,P_1992,P_2015} by linear quantum stochastic differential equations (QSDEs) driven by quantum Wiener processes representing external bosonic fields.
Linear QSDEs form a wide class of tractable models of open quantum dynamics, which are studied in linear quantum systems theory \cite{NY_2017,P_2017}.  The predictions, obtained in the framework of this theory, employ certain assumptions on the coefficients of the QSDEs and the system-field density operators which may differ from their nominal models. The accuracy of modelling such 
systems and networks \cite{GJ_2009,JG_2010} is important for their applications in modern quantum technologies \cite{NC_2000,WM_2008}.

In addition to the $\cH_\infty$ and guaranteed cost LQG settings \cite{JNP_2008,SPJ_2007},  the issue of robustness with respect to unmodelled dynamics and quantum statistical uncertainties is addressed, for example, in quantum risk-sensitive control  and filtering. This approach employs   time-ordered exponentials \cite{J_2004,J_2005} or  the usual  operator  exponentials \cite{B_1996,YB_2009,VPJ_2018a} which give rise to higher-order (exponential)  moments of relevant quantum variables as cost functionals to be minimized. In particular, the quadratic-exponential functional (QEF) \cite{VPJ_2018a} is organized as the exponential moment of the integral of a  quadratic form of the quantum system variables over a bounded time interval,
similarly to its predecessors in classical risk-sensitive control \cite{BV_1985,J_1973,W_1981}. The minimization of the QEF improves the upper bound \cite{VPJ_2018b}  for the worst-case mean square costs in the presence of quantum statistical uncertainty,  described in terms of the quantum relative entropy  \cite{NC_2000,OW_2010} of the actual system-field state with respect to its nominal model. A similar role is played by the QEF minimization for the Cramer type tail distribution bounds \cite{VPJ_2018a} for the quantum trajectories.

The fact that the QEF minimization makes the open quantum system dynamics more robust and conservative is the main motivation for the development of methods for its practical computation (which is different from that of the time-ordered exponential counterpart in \cite{DDJW_2006,J_2004,J_2005}).
To this end, Lie-algebraic techniques \cite{VPJ_2019a}, parametric randomization \cite{VPJ_2018c}, quantum Karhunen-Loeve  expansions \cite{VPJ_2019b,VJP_2019}, and a Girsanov type representation \cite{VPJ_2019c}  have recently been proposed. These results have led to  a frequency-domain formula \cite{VPJ_2019d} for the infinite-horizon  asymptotic growth rate of the QEF for invariant Gaussian states \cite{KRP_2010} of stable OQHOs driven by vacuum fields.

The present paper uses the QEF growth rate as a robust performance criterion for   a quantum risk-sensitive     control problem. This setting is concerned with a measurement-based  feedback connection of a quantum plant with a
classical controller. The plant is modelled as a multimode OQHO with a multichannel  quantum Wiener process at the input. Its dynamics are also affected by the output of the  controller, which is a linear time invariant system governed by a classical SDE driven by the observation process. The control objective is to stabilize the closed-loop system and minimize the QEF growth rate,  which penalizes the plant variables and the control signal. We combine the frequency-domain representation of the QEF growth rate \cite{VPJ_2019d}, mentioned above, with variational techniques \cite{VP_2013a,VP_2013b} in order to compute the Frechet derivatives of the cost functional with respect to the controller matrices. This leads to a set of first-order necessary conditions of optimality for 
the controller and allows a gradient descent algorithm, similar to \cite{SVP_2017},  to be outlined for the numerical solution of the control problem.

The paper is organized as follows.
Section~\ref{sec:sys} describes the class of  linear quantum plants and the general structure of measurement-based feedback.
Section~\ref{sec:meas} specifies nondemolition measurements and classical linear controllers being considered.
Section~\ref{sec:QEF} provides the frequency-domain representation for the QEF growth rate as a performance criterion for the closed-loop system.
Section~\ref{sec:opt} establishes first-order necessary conditions of optimality for the resulting risk-sensitive  quantum  control problem and outlines its numerical solution via gradient descent.
Section~\ref{sec:conc} makes concluding remarks. Appendix~\ref{sec:matfun} provides subsidiary lemmas on differentiation of functions of matrices.

\section{Quantum plant with measurement-based feedback}
\label{sec:sys}

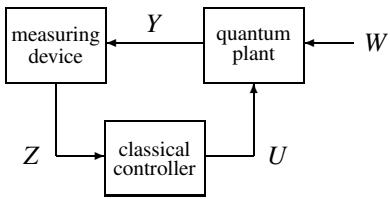
\begin{figure}[htbp]
\unitlength=0.65mm
\linethickness{0.4pt}
\qquad\begin{picture}(70.00,45.00)
    \put(60,25){\framebox(20,15)[cc]{{}}}
    \put(60,27){\makebox(20,15)[cc]{{\small quantum}}}
    \put(60,23){\makebox(20,15)[cc]{{\small plant}}}

    \put(20,25){\framebox(20,15)[cc]{{}}}
    \put(20,27){\makebox(20,15)[cc]{{\small measuring}}}
    \put(20,23){\makebox(20,15)[cc]{{\small device}}}

    \put(95,33){\makebox(0,0)[cc]{$W$}}
    \put(50,37){\makebox(0,0)[cc]{$Y$}}
    \put(75,10){\makebox(0,0)[cc]{$U$}}
    \put(25,10){\makebox(0,0)[cc]{$Z$}}

    \put(90,33){\vector(-1,0){10}}

    \put(60,33){\vector(-1,0){20}}

    \put(30,25){\line(0,-1){15}}
    \put(30,10){\vector(1,0){10}}
    \put(60,10){\line(1,0){10}}
    \put(70,10){\vector(0,1){15}}

    \put(40,2){\framebox(20,15)[cc]{{}}}
    \put(40,4){\makebox(20,15)[cc]{{\small classical}}}
    \put(40,0){\makebox(20,15)[cc]{{\small controller}}}
\end{picture}\vskip-2mm
\caption{A block diagram of the quantum plant (organized as an OQHO)
with an output field $Y$, driven by the quantum Wiener process $W$  and the classical actuator signal $U$ of a classical controller based on the measurement signal $Z$.}
\label{fig:OQHO}
\end{figure}

We consider a quantum plant,  which interacts with the environment, including the external bosonic fields 
and a measurement-based feedback loop.  The feedback is carried out by a classical controller  (for example, using digital electronics) which processes the output $Z$ of a measuring device and produces an actuator signal $U$ affecting the quantum plant dynamics; see Fig.~\ref{fig:OQHO}.
The plant is organized as a linear quantum stochastic system (an OQHO) and is endowed with an even number $n$ of dynamic variables $X_1, \ldots, X_n$ (for example, pairs of conjugate  quantum mechanical positions and momenta \cite{S_1994}). These plant variables  are time-varying self-adjoint operators on a complex separable Hilbert space $\fH$ (or a dense domain thereof), which are assembled into a vector $X:=(X_k)_{1\< k\< n}$ 
(vectors are organized as columns unless specified otherwise, and the time arguments are often omitted for brevity).
The plant variables satisfy the Weyl canonical commutation relations (CCRs) \cite{F_1989} whose infinitesimal Heisenberg form is given by
\begin{equation}
\label{Theta}
    [X, X^{\rT}]
    :=    ([X_j,X_k])_{1\< j,k\< n}
       =    XX^{\rT}- (XX^{\rT})^{\rT}
       =
     2i \Theta, 
\end{equation}
where $[\alpha,\beta]:= \alpha \beta-\beta\alpha$ is the commutator of linear operators, $i:= \sqrt{-1}$ is the imaginary unit, and $\Theta$ 
is a 
real antisymmetric matrix. 
The Heisenberg dynamics of the plant variables are governed by a linear QSDE
\begin{equation}
\label{dX}
    \rd X
    =
    (A X +  E U) \rd t+ B \rd W,
\end{equation}
where $A\in \mR^{n\x n}$, $B \in \mR^{n\x m}$, $E\in \mR^{n\x d}$  are constant matrices.  The ingredients of this QSDE are described below.
The diffusion term $B \rd W$ in (\ref{dX}) involves the forward Ito increment
of the vector $W:=(W_k)_{1\< k\< m}$ 
of an even number $m$ of quantum Wiener processes $W_1, \ldots, W_m$ which are time-varying self-adjoint operators on a symmetric Fock space $\fF$ \cite{P_1992,PS_1972}. These operators  represent the input bosonic fields and have a complex positive semi-definite Hermitian Ito matrix $\Omega \in \mC^{m\x m}$:
\begin{equation}
\label{WW}
    \rd W \rd W^{\rT}
    =
    \Omega \rd t,
    \qquad
    \Omega := I_m + iJ,
\end{equation}
where $I_m$ is the identity matrix of order $m$. Its imaginary part $J$, given by
\begin{equation}
\label{J}
    J
        :=
        \Im \Omega
        =
        \bJ \ox I_{m/2}
        =
       {\begin{bmatrix}
           0 & I_{m/2}\\
           -I_{m/2} & 0
       \end{bmatrix}},
       \qquad
       \bJ
       :=
       {\begin{bmatrix}
         0 & 1\\
         -1 & 0
       \end{bmatrix}},
\end{equation}
is an orthogonal real antisymmetric matrix of order $m$ (and hence, $J^2=-I_m$). This matrix specifies the two-point CCRs for the quantum Wiener processes as
\begin{equation}
\label{WWcomm}
    [W(s), W(t)^\rT]
    =
    2i\min(s,t) J,
    \qquad
    s,t\>0,
\end{equation}
which, in the case $s=t$, leads to $[W(t),W(t)^\rT] = 2i t J$, and hence,
\begin{equation}
\label{dWdWcomm}
    [\rd W, \rd W^\rT] = \rd [W, W^\rT] - [\rd W, W^\rT]-[W, \rd W^\rT]=
     2iJ\rd t
\end{equation}
in view of the commutativity between the Ito increments of the quantum Wiener process and adapted quantum processes. The adaptedness is understood in the sense of the
filtration $(\fH_t)_{t\>0}$ given by
\begin{equation}
\label{fHt}
    \fH_t
    :=
    \fH_0 \ox \fF_t,
\end{equation}
where $\fH_0$ is a complex separable Hilbert space for the action of the initial plant variables $X_1(0), \ldots, X_n(0)$, and  $\fF_t$ is the Fock subspace associated with the time interval $[0,t]$,  with the tensor product
\begin{equation}
\label{fH}
    \fH := \fH_0 \ox \fF
\end{equation}
specifying the plant-field Hilbert space. The drift of the QSDE (\ref{dX}) involves the classical actuator signal $U:=(U_k)_{1\< k\< d}$ with values in $\mR^d$,
which is the output of the controller satisfying the nondemolition condition \cite{B_1983} with respect to the plant variables:
\begin{equation}
\label{XUcomm}
  [X(t), U(s)^\rT]
  =
  0,
  \qquad
  t\> s\> 0.
\end{equation}
The energetics of the plant as an OQHO is captured by the following Hamiltonian $H$ and vector $L$  of plant-field coupling operators which are quadratic and linear functions of the plant variables, respectively:
\begin{equation}
\label{HL}
    H
    :=
    \tfrac{1}{2}
    X^\rT R X
    +
    U^\rT N X,
    \qquad
    L:=
    (L_k)_{1\< k\< m}
    =
    MX.
\end{equation}
Here, $R = R^{\rT} \in \mR^{n\x n}$ is the energy matrix, and $M \in \mR^{m\x n}$ and $N\in \mR^{d\x n}$ are the matrices of coupling of the plant to the external bosonic fields and the controller. In view of the commutativity (\ref{XUcomm}),  the control signal $U$ plays the role of a time-varying  classical parameter of the Hamiltonian $H$ in (\ref{HL}), thus  influencing the energy of the quantum plant. Assuming the identity scattering matrix\footnote{when there is no photon exchange between the field channels, which effectively removes the gauge processes from consideration} \cite{HP_1984,P_1992} for what follows,
the energy operators in (\ref{HL}) specify the
stochastic Schr\"{o}dinger equation
\begin{equation}
\label{dV}
    \rd V
    = -V
    \big(i(H\rd t + L^{\rT} \rd W) + \tfrac{1}{2}L^{\rT}\Omega L\rd t\big)
\end{equation}
for a time-varying unitary operator $V$ on the plant-field space (\ref{fH}), with the identity operator  on $\fH$ as the initial condition $V(0) = \cI_{\fH}$.  The operator $V$ encodes the controlled plant-field interaction and  gives rise to the quantum stochastic flow of the plant variables:
\begin{equation}
\label{Xflow}
    X(t) = V(t)^{\dagger}(X(0)\ox \cI_{\fF})V(t),
\end{equation}
with $(\cdot)^\dagger$ the operator adjoint.
The QSDE (\ref{dX}) can be obtained from (\ref{Xflow}) by using (\ref{Theta}),  (\ref{XUcomm})--(\ref{dV}) and the commutativity between the forward Ito increments of the quantum Wiener process and adapted quantum processes with respect to the filtration (\ref{fHt}) mentioned above, whereby $[\rd W(t), X(s)^\rT] = 0$ for all $t\>s\>0$. The matrices $A$, $B$, $E$ in (\ref{dX}) are parameterized by the energy matrix $R$ and the plant-field and plant-controller coupling matrices $M$, $N$ in (\ref{HL}) as
\begin{equation}
\label{ABE}
  A = 2\Theta (R + M^\rT J M),
  \qquad
  B = 2\Theta M^{\rT},
  \qquad
  E = 2\Theta N^\rT.
\end{equation}
Due to this structure,  the matrices $A$, $B$ satisfy the algebraic equation
\begin{equation}
\label{ABPR}
  A \Theta + \Theta A^{\rT} + BJB^{\rT} = 0,
\end{equation}
which is one of the physical realizability (PR) conditions \cite{JNP_2008,SP_2012} and is closely related to the preservation of the CCRs (\ref{Theta}) in time. This property is a corollary of the flow (\ref{Xflow}) being a unitary similarity transformation of quantum variables on $\fH$.
The action of the flow on the input fields $W$ produces the vector
\begin{equation}
\label{Yflow}
    Y(t)
    :=
    (Y_k(t))_{1\< k\< m}
     =
 V(t)^{\dagger}(\cI_{\fH_0}\ox W(t))V(t)
\end{equation}
of output fields, which are time-varying self-adjoint quantum variables on the plant-field space (\ref{fH}) satisfying the QSDE
\begin{equation}
\label{dY}
  \rd Y = 2J L \rd t + \rd W = 2J MX \rd t + \rd W,
\end{equation}
with $L$ given by (\ref{HL}).
The flow in (\ref{Xflow}) and (\ref{Yflow}) preserves the commutativity between the plant and output field variables (which commute initially  as operators on different spaces $\fH_0$ and $\fF$):
\begin{equation}
\label{XYcomm}
        [X(t),Y(t)^{\rT}]
     =
     V(t)^\dagger [X(0) \ox \cI_{\fF},\cI_{\fH_0}\ox W(0)^\rT] V(t)
      =0
\end{equation}
for all $t\>0$. 
However, the plant output fields $Y_1, \ldots, Y_m$ do not commute with each other since $[Y(t), Y(t)^{\rT}] = [W(t), W(t)^{\rT}] = 2it J$ in view of (\ref{WWcomm}), which was used in (\ref{dWdWcomm}). The noncommutativity of the output fields  makes them inaccessible to simultaneous measurement.

\section{Nondemolition measurements and linear controllers}
\label{sec:meas}

 The measuring device in Fig.~\ref{fig:OQHO} converts the plant output $Y$ to a  multichannel observation process $Z:=(Z_k)_{1\< k\< r}$ 
consisting of $r$ time-varying self-adjoint quantum variables, which commute with future plant variables and between themselves at all times:
\begin{equation}
\label{XZZcomm}
    [X(t), Z(s)^{\rT}] = 0,
    \qquad
    [Z(t), Z(s)^{\rT}] = 0,
    \qquad
    t\> s\> 0,
\end{equation}
where the first equality is closely related to the nondemolition condition (\ref{XUcomm}).
For any time $t\>0$, any operator $f(X(t))$ (an appropriate operator-valued extension of a  complex-valued function $f$ applied to the plant variables) and the past observation history
\begin{equation}
\label{Zt}
    \fZ_t
    :=
    \{
        Z_1(s), \ldots, Z_r(s):\ 0\<s\< t
    \}
\end{equation}
form a set of pairwise commuting (and hence, compatible) quantum variables. 
The commutative von Neumann algebra $\cZ_t$,  generated by the past observation history $\fZ_t$  from (\ref{Zt}), is the information available to the causal controller at the moment of time $t\>0$. The corresponding actuator signal $U(t)$ is a $\cZ_t$-adapted $\mR^d$-valued vector (so that there is a deterministic function of the observation history which selects $U(t)$ as a particular element of the algebra $\cZ_t$).
As a nondemolition measurement model in the sense of (\ref{XZZcomm}), we will use (following \cite{N_2014}) the static relation
\begin{equation}
\label{ZY}
    Z
     = D Y.
\end{equation}
Here, $D\in \mR^{r\x m}$ is a matrix with $r\< \frac{m}{2}$ rows satisfying the conditions
\begin{equation}
\label{DD_DJD}
  DD^{\rT}  \succ 0,
  \qquad
  DJD^{\rT}  = 0,
\end{equation}
the first of which is equivalent to $D$ being of full row rank. From (\ref{dY}), it follows that the process $Z$ in (\ref{ZY}) is governed by the QSDE
\begin{equation}
\label{dZ}
  \rd Z = CX \rd t + D \rd W,
\end{equation}
where the matrix $C\in \mR^{r\x n}$ is related to the plant-field coupling matrix $M$ from (\ref{HL}) as
\begin{equation}
\label{C}
  C:= 2DJM
\end{equation}
and satisfies another PR condition which complements (\ref{ABPR}):
\begin{equation}
\label{CPR}
  \Theta C^\rT + BJD^\rT = 0.
\end{equation}
This property pertains to the commutativity $[X,Z^\rT] = [X,Y^\rT]D^\rT\\ = 0$, which the process $Z$ in (\ref{ZY}) inherits from the plant output $Y$ in (\ref{XYcomm}).  Furthermore,
in view of (\ref{dWdWcomm}), the second of the conditions (\ref{DD_DJD}) implies that $[\rd Z, \rd Z^{\rT}] = D[\rd W, \rd W^\rT]D^\rT = 2iD JD^{\rT}\rd t=0$,  which indeed makes the process $Z$ self-commuting and its entries $Z_1, \ldots, Z_r$ accessible to simultaneous continuous measurement. Up to an isomorphism, $Z$ is a classical  Ito process \cite{KS_1991} with values in  $\mR^r$ and a positive definite diffusion matrix $D\Omega  D^\rT =DD^{\rT}+iDJD^{\rT}=DD^{\rT}$ in view of (\ref{WW}), (\ref{DD_DJD}), so that 
 $ \rd Z \rd Z^\rT = D \rd W \rd W^\rT D^\rT = D\Omega D^\rT \rd t = DD^\rT \rd t$. 
The dynamics of the closed-loop system with the measurement-based feedback are specified by the QSDEs (\ref{dX}), (\ref{dZ}) in combination with a particular causal control law. A linear dynamic feedback is provided by
\begin{equation}
\label{dxi}
  \rd \xi  = a \xi \rd t + b \rd Z,
  \qquad
  U  = c \xi,
\end{equation}
where the controller state $\xi$ is a classical Ito process in $\mR^n$   driven by the observation process $Z$ according to a linear SDE, with the initial condition $\xi(0)$ being $\fH_0$-adapted. 
The controller dynamics are specified by
constant matrices $a\in \mR^{n\x n}$, $b \in \mR^{n\x r}$, $c \in \mR^{d\x n}$ which are not constrained by quantum PR conditions in contrast to the plant matrices $A$, $B$, $C$, $D$  in (\ref{ABE}), (\ref{ABPR}), (\ref{C}), (\ref{CPR}).
The resulting closed-loop system is endowed with an augmented vector
\begin{equation}
\label{cX}
  \cX
  :=
  {\small\begin{bmatrix}
    X\\
    \xi
  \end{bmatrix}}
\end{equation}
of the dynamic variables of the quantum plant (\ref{dX}) and the classical controller (\ref{dxi}) and is governed by the linear QSDE
\begin{equation}
\label{dcX}
  \rd \cX = \cA \cX \rd t + \cB \rd W,
\end{equation}
where the matrices $\cA \in \mR^{2n\x 2n}$, $\cB \in \mR^{2n\x m}$ are computed as
\begin{equation}
\label{cAB}
  \cA
  :=
  {\begin{bmatrix}
    A & Ec\\
    bC & a
  \end{bmatrix}},
  \qquad
  \cB
  :=
  {\begin{bmatrix}
    B\\
    bD
  \end{bmatrix}}
\end{equation}
and depend affinely on the controller matrices $a$, $b$, $c$.
From (\ref{ABPR}), (\ref{CPR}) and the second equality in (\ref{DD_DJD}), it follows  that the matrices (\ref{cAB}) satisfy a similar PR  condition
\begin{align}
\nonumber
  \cA \Gamma & + \Gamma \cA^\rT + \cB J \cB^\rT\\
\label{cABPR}
  & =
  {\begin{bmatrix}
    A \Theta + \Theta A^{\rT} + BJB^{\rT} & (\Theta C^\rT + BJD^\rT)b^\rT \\
    b(C \Theta  + DJB^\rT) & bDJD^\rT b^\rT
  \end{bmatrix}}=0,
\end{align}
where $\Gamma$ is the joint CCR matrix of the plant and controller variables in (\ref{cX}):
\begin{equation}
\label{bTheta}
    [\cX,\cX^\rT]
    =
  {\begin{bmatrix}
    [X,X^\rT] & [X,\xi^\rT]\\
    [\xi, X^\rT] & [\xi,\xi^\rT]
  \end{bmatrix}}
  =
    2i\Gamma,
    \qquad
  \Gamma
  :=
  {\begin{bmatrix}
    \Theta & 0\\
    0 & 0
  \end{bmatrix}}
\end{equation}
in accordance with (\ref{Theta}). Note that (\ref{cABPR}) holds for any controller matrices $a$, $b$, $c$ in (\ref{dxi}) regardless of the particular structure  of the matrix  $E$ in (\ref{ABE}) in terms of the plant-controller coupling matrix $N$. As a solution of the linear QSDE (\ref{dcX}),  the process (\ref{cX}) satisfies
$    \cX(t) = \re^{(t-s)\cA} \cX(s) + \int_s^t \re^{(t-\tau) \cA}\cB \rd W(\tau)$ for
all $    t\> s\> 0$,
which leads 
to the two-point CCRs
\begin{equation}
\label{cXcX}
    [\cX(s), \cX(t)^\rT]
    =
    2i\Lambda(s-t),
    \qquad
    s,t\>0,
\end{equation}
with
\begin{equation}
\label{Lambda}
    \Lambda(\tau)
    :=
    \left\{
    {\begin{matrix}
    \re^{\tau \cA}\Gamma & {\rm if}\  \tau\> 0\\
    \Gamma \re^{-\tau \cA^\rT}& {\rm if}\  \tau< 0
    \end{matrix}}
    \right.
    =
    {\begin{bmatrix}
      \Lambda_{11}(\tau) & \Lambda_{12}(\tau)\\
      \Lambda_{21}(\tau) & 0
    \end{bmatrix}}
    =
    -\Lambda(-\tau)^\rT.
\end{equation}
The one-point CCRs (\ref{bTheta}) are a particular case of (\ref{cXcX}), (\ref{Lambda}) since $\Lambda(0) = \Gamma$. By the last equality in (\ref{Lambda}), the $(n\x n)$-blocks $\Lambda_{jk}$   satisfy 
$    \Lambda_{11}(\tau) = -\Lambda_{11}(-\tau)^\rT$ and
$
    \Lambda_{12}(\tau) = -\Lambda_{21}(-\tau)^\rT$
for all $
    \tau \in \mR$. 
Furthermore,
\begin{equation}
\label{L12L21}
    \Lambda_{12}(\tau)=0,
    \qquad
    \Lambda_{21}(-\tau) = 0,
    \qquad
    \tau\>0,
\end{equation}
and the functions $\Lambda_{11}$, $\Lambda_{21}$, considered on $\mR_+$, satisfy  the ODEs
\begin{equation}
\label{L11dot_L21dot}
    \Lambda_{11}^{^\centerdot}
    =
    A \Lambda_{11} + Ec \Lambda_{21},
    \qquad
    \Lambda_{21}^{^\centerdot}
     =
    bC \Lambda_{11} + a \Lambda_{21},
\end{equation}
with the initial conditions $\Lambda_{11}(0) = \Theta$, $\Lambda_{21}(0) = 0$, where $\dot{(\ )}$ is the time derivative.  Indeed, (\ref{L12L21}) follows from the sparsity of the plant-controller CCR matrix $\Gamma$ in (\ref{bTheta}), while the ODEs (\ref{L11dot_L21dot}) are obtained from the blockwise form of the ODE $\Lambda^{^\centerdot}(\tau) = \cA \Lambda(\tau)$ for all  $\tau\>0$ in view of (\ref{Lambda}) and the structure of the matrix $\cA$ in (\ref{cAB}), with the initial conditions coming from the first block-column of the matrix $\Gamma$.

Although the one-point CCR matrix $\Gamma$ in (\ref{bTheta}) does not depend on the controller matrices $a$, $b$, $c$, the latter affect the two-point CCRs of the plant and controller variables. In particular, the second ODE in (\ref{L11dot_L21dot}) implies that $\Lambda_{21}^{^\centerdot}(0+)
     =
    bC \Lambda_{11}(0) + a \Lambda_{21}(0) = bC\Theta$ depends on $b$. Moreover, the two-point CCR kernel $\Lambda_{11}$ of the plant variables in (\ref{Lambda}) satisfies $\ddot{\Lambda}_{11}(0+) = (\cA^2)_{11}\Theta = (A^2 + EcbC)\Theta$ and is also affected by the controller matrices.

\section{Quadratic-exponential performance criterion}
\label{sec:QEF}

Consider a quantum risk-sensitive setting, which describes
performance of the measurement-based feedback over the time interval $[0,T]$ (for a given time horizon $T>0$) in terms of the QEF
\begin{equation}
\label{Xi}
    \Xi_T
    :=
    \bE \re^{\frac{\theta}{2} Q_T},
\end{equation}
where $\bE \zeta := \Tr(\rho \zeta)$ is the quantum expectation over an underlying density operator $\rho$ on the plant-field space $\fH$ in (\ref{fH}). Here, $\theta>0$ is a given risk sensitivity parameter (the dependence on $\theta$ is often omitted for brevity)  which specifies the exponential penalty on the positive semi-definite self-adjoint quantum variable
\begin{equation}
\label{Q}
    Q_T
    :=
    \int_0^T
    \cV(t)^\rT \cV(t)
    \rd t
    =
    \int_0^T
    \cX(t)^\rT \cC^\rT \cC\cX(t)
    \rd t
\end{equation}
on the subspace $\fH_T$  in (\ref{fHt}). The variable $Q_T$
 depends quadratically on the history of the plant and controller variables in (\ref{dX}), (\ref{dxi}), (\ref{cX}) over the time interval $[0,T]$ through an auxiliary process
 \begin{equation}
\label{cV}
  \cV
  :=
  S X + K U = \cC \cX,
\end{equation}
which consists of $\nu$ time-varying self-adjoint quantum variables, where
$S \in \mR^{\nu \x n}$, $K \in \mR^{\nu \x d}$ are given weighting matrices. Accordingly,
\begin{equation}
\label{cC}
  \cC
  :=
  {\begin{bmatrix}
    S & &Kc
  \end{bmatrix}}
\end{equation}
is a $(\nu\x 2n)$-matrix which depends affinely on the controller matrix $c$ from (\ref{dxi}).

The above setting can capture the conventional risk-sensitive control problem by letting
$S := {\small\begin{bmatrix}
  \sqrt{\Pi_1}\\
  0
\end{bmatrix}}$ and $K := {\small\begin{bmatrix}
  0 \\
  \sqrt{\Pi_2}
\end{bmatrix}}$, where  $\Pi_1 \in \mR^{n\x n}$, $\Pi_2\in \mR^{d\x d}$ are positive definite symmetric matrices. In this case, (\ref{Q}) takes the form $Q_T= \int_0^T (X(t)^\rT \Pi_1 X(t) + \|U(t)\|_{\Pi_2}^2)\rd t$, and (\ref{Xi}) penalizes both the plant variables and the controller output variables, with $\|u\|_\Pi: = \sqrt{\Tr(u^\rT \Pi u)} = |\sqrt{\Pi}u|$ a vector norm generated by a positive definite matrix $\Pi$. Alternatively,  a quantum risk-sensitive filtering problem is  obtained in the case $E=0$ (when the plant dynamics (\ref{dX}) are not affected by the controller)  by letting $S: = \sqrt{\Pi}$ and $K:= -\sqrt{\Pi}$, where $\Pi=\Pi^\rT \in \mR^{n\x n}$ is a positive  semi-definite matrix. In this case, $d=n$, and  (\ref{Q}) acquires the form   $Q_T=\int_0^T (X(t)-U(t))^\rT \Pi (X(t)-U(t))\rd t$. Accordingly,  the controller becomes an observer, with its output $U$ playing the role of an estimator for the plant variables, so that (\ref{Xi}) penalizes the ``estimation error'', cf. \cite{YB_2009}.

We will now return to the general setting (\ref{cV}) with arbitrary weighting matrices $S$, $K$.  In view of (\ref{cXcX}), (\ref{Lambda}),   the process $\cV$ satisfies the two-point CCRs
\begin{equation}
\label{cVcVcomm}
    [\cV(s), \cV(t)^\rT]
    =
    2i\mho(s-t),
    \qquad
    s,t\>0,
\end{equation}
with
\begin{equation}
\label{mho}
  \mho(\tau):= \cC \Lambda(\tau) \cC^\rT = -\mho(-\tau)^\rT,
  \qquad
  \tau \in \mR,
\end{equation}
where the last equality is inherited from that in (\ref{Lambda}). The two-point CCR function $\mho$ in (\ref{mho}) specifies a skew self-adjoint integral operator $\cL_T$  on the Hilbert space $L^2([0,T],\mC^\nu)$ of square integrable $\mC^\nu$-valued functions on the time interval $[0,T]$ as
\begin{equation}
\label{cL}
    \cL_T(f)(s)
    :=
    \int_0^T
    \mho(s-t) f(t)
    \rd t,
    \qquad
    0\< s \< T.
\end{equation}
While the commutation structure of the plant and controller variables (and also (\ref{cVcVcomm})) does not depend on the plant-field quantum state, we will be concerned 
with the case of vacuum input fields and \emph{stabilizing} controllers which make the matrix $\cA$ in (\ref{cAB}) Hurwitz.
An appropriate modification of the results of \cite{VPJ_2018a} shows that in this case the closed-loop system variables have a unique  invariant multipoint zero-mean Gaussian quantum state. This property is inherited by the process $\cV$ in (\ref{cV}). The corresponding  two-point quantum covariance function
\begin{equation}
\label{EVV}
\bE(\cV(s)\cV(t)^\rT)
  =
  P(s-t) + i\mho(s-t),
  \qquad
  s,t\> 0
\end{equation}
has the imaginary part (\ref{mho}) (regardless of the quantum state), and the real part 
\begin{equation}
\label{P}
    P(\tau)
    =
    \left\{
    {\begin{matrix}
    \cC\re^{\tau \cA}\Sigma \cC^\rT& {\rm if}\  \tau\> 0\\
    \cC \Sigma \re^{-\tau \cA^\rT}\cC^\rT & {\rm if}\  \tau < 0
    \end{matrix}}
    \right.
    =
    P(-\tau)^\rT,
    \qquad
    \tau \in \mR.
\end{equation}
Here, $\Sigma:= \Re \bE (\cX\cX^\rT)$ is a real positive semi-definite symmetric matrix of order $2n$, which describes the one-point covariances of the plant and controller variables and satisfies the algebraic Lyapunov equation (ALE)
  $\cA \Sigma + \Sigma\cA^\rT + \cB\cB^\rT=0$, whose structure is similar to (\ref{cABPR}).  The kernel (\ref{P}) gives rise to a positive semi-definite self-adjoint integral  operator $\cP_T$ on $L^2([0,T], \mC^\nu)$ as
\begin{equation}
\label{cP}
    \cP_T(f)(s)
    :=
    \int_0^T P(s-t)f(t)\rd t,
    \qquad
    0\< s \< T.
\end{equation}
Moreover, the property $\cP_T \succcurlyeq 0$ is a corollary of positive semi-definiteness of the self-adjoint operator $\cP_T+i\cL_T$ on $L^2([0,T], \mC^\nu)$. Application of appropriately modified results of \cite{VPJ_2019c} to the process $\cV$,  associated by  (\ref{cV}) with the closed-loop system in the invariant multipoint Gaussian quantum state, allows the QEF (\ref{Xi}) to be  computed as
\begin{equation}
\label{lnXi}
  \ln \Xi_T
    =
    -  \tfrac{1}{2}
  \Tr (\ln\cos (\theta\cL_T) + \ln (\cI - \theta \cP_T\cK_T )),
\end{equation}
provided the operator $\cL_T$ in (\ref{cL}) has no zero eigenvalues.
Here,
\begin{equation}
\label{cK}
    \cK_T
    :=
    \tanhc(i\theta \cL_T) = \tanc (\theta\cL_T)
\end{equation}
is a positive definite self-adjoint operator on $L^2([0,T],\mC^\nu)$, and $\tanhc z := \tanc (-iz)$ is a hyperbolic version of the function $\tanc z := \frac{\tan z}{z}$ extended as $\tanc 0:=1$ by continuity. Note that the operator $\cK_T$ is nonexpanding in the sense that $\cK_T \preccurlyeq \cI$, where $\cI$ is the identity operator on  $L^2([0,T],\mC^\nu)$. With $\cP_T\cK_T$ being a compact operator (which is isospectral to the positive semi-definite self-adjoint operator $\sqrt{\cK_T} \cP_T \sqrt{\cK_T}$ associated with (\ref{cP}), (\ref{cK})), the representation (\ref{lnXi}) is valid under the condition
\begin{equation}
\label{spec}
    \theta \lambda_{\max}(\cP_T\cK_T) < 1,
\end{equation}
where $\lambda_{\max}(\cdot)$ is the largest eigenvalue. Similarly to \cite{VPJ_2019c}, (\ref{lnXi}) is obtained by using a quantum Karhunen-Loeve expansion \cite{VPJ_2019b,VJP_2019} of the process $\cV$ in (\ref{cV}) over an orthonormal eigenbasis of the operator $\cL_T$ in (\ref{cL}), where the condition that $\cL_T$ has no zero eigenvalues plays its part. A formulation of this condition in terms of the state-space matrices of the system  will be discussed elsewhere in view of $\cC$ in (\ref{cV}) being a nonsquare matrix. Assuming that the operator $\cL_T$ has no zero eigenvalues for all sufficiently large $T>0$, application of \cite[Theorem~1]{VPJ_2019d} to the QEF $\Xi_T$ in (\ref{Xi})--(\ref{cC}) leads to the following infinite-horizon growth rate for its logarithm (\ref{lnXi}):
\begin{equation}
\label{Ups}
    \Ups
    :=
\lim_{T\to +\infty}
    \big(
        \tfrac{1}{T}
        \ln \Xi_T
    \big)
     =
    -
    \tfrac{1}{4\pi}
    \int_{\mR}
    \ln\det
    \Delta(\lambda)
    \rd \lambda.
\end{equation}
Here,
\begin{align}
\nonumber
    \Delta(\lambda)
    & :=
    \cos(
        \theta \Psi(\lambda)
    ) -
        \theta
        \Phi(\lambda)
        \sinc
        (\theta \Psi(\lambda))\\
\label{D}
    & =
    \cos(
        \theta \Psi(\lambda)
    ) -
        \Phi(\lambda)
        \Psi(\lambda)^{-1}
        \sin
        (\theta \Psi(\lambda))
\end{align}
(with $\sinc z := \frac{\sin z}{z}$ extended  as $\sinc 0 := 1$ by continuity) is defined in terms of the Fourier transforms of the covariance and commutator kernels (\ref{P}), (\ref{mho}):
\begin{align}
\label{Phi0}
    \Phi(\lambda)
    & :=
    \int_\mR \re^{-i\lambda t }
    P(t)
    \rd t
    =
    F(i\lambda) F(i\lambda)^*,\\
\label{Psi0}
    \Psi(\lambda)
    & :=
    \int_\mR \re^{-i\lambda t }
    \mho(t)
    \rd t
    =
    F(i\lambda) J F(i\lambda)^*,
    \qquad
    \lambda \in \mR,
\end{align}
see also \cite[Eq.~(5.8)]{VPJ_2019a}, where $(\cdot)^*:= {{\overline{(\cdot)}}}^\rT$ is the complex conjugate transpose. Also,
\begin{equation}
\label{F0}
    F(s)
    :=
    \cC
    G(s)
    \cB,
    \qquad
    s \in \mC,
\end{equation}
is the transfer function for the closed-loop system (\ref{dcX})  from the incremented input quantum Wiener process $W$ to the process $\cV$ in (\ref{cV}), with
\begin{equation}
\label{G}
  G(s):= (sI_{2n} - \cA)^{-1}.
\end{equation}
In order for the relation (\ref{Ups}) to be valid, the risk sensitivity parameter $\theta>0$ in (\ref{Xi}) has to be small enough in the sense that
\begin{equation}
\label{spec1}
    \theta
    \sup_{\lambda \in \mR}
    \lambda_{\max}
    (
        \Phi(\lambda)
        \tanc
        (\theta \Psi(\lambda))
    )
    < 1,
\end{equation}
which originates from (\ref{spec}). Note that $\Phi(\lambda)$ is a complex positive semi-definite Hermitian  matrix, while  $\Psi(\lambda)$  is skew Hermitian at any frequency $\lambda \in \mR$. The function $\Phi+i\Psi$ is the Fourier transform  of the quantum covariance kernel $P+i\mho$ from (\ref{EVV}), thus playing the role of a ``quantum spectral density'' for the process $\cV$ in (\ref{cV}).

As a function of $\theta$ (subject to (\ref{spec1})), the QEF growth rate (\ref{Ups}) can be used for quantifying the large deviations of quantum trajectories \cite{VPJ_2018a} of the closed-loop system and its robustness with respect to state uncertainties in terms of quantum relative entropy \cite{OW_2010}  (see \cite[Section~IV]{VPJ_2018b}
 and references therein). These quantum robustness bounds are similar to those in minimax LQG control of classical stochastic systems \cite{DJP_2000,P_2006,PJD_2000} and depend on the QEF growth rate $\Ups$ in a monotonic fashion.  
This leads to a quantum risk-sensitive optimal control problem
\begin{equation}
\label{opt}
  \Ups \longrightarrow \inf,
\end{equation}
where the QEF growth rate (\ref{Ups}) is minimized for a given $\theta>0$ over the state-space matrices $a$, $b$, $c$  of stabilizing controllers (\ref{dxi}) subject to the constraint (\ref{spec1}). 
 In comparison with classical risk-sensitive control \cite{BV_1985,J_1973,W_1981}, the controller matrices influence the cost functional (\ref{Ups}) not only through the statistical properties of the plant and controller variables, captured by the function $\Phi$ in (\ref{Phi0}), but also through their commutation  structure described by $\Psi$ in (\ref{Psi0}), which enters (\ref{Ups}) in view of (\ref{D}).

\section{First-order necessary conditions of optimality}
\label{sec:opt}

Due to the closed-loop system matrix $\cA$ being Hurwitz for stabilizing controllers, the cost functional (\ref{Ups}) is a smooth function of the state-space matrices $a$, $b$, $c$ of such controllers subject to (\ref{spec1}). Similarly to the variational approach to quantum control and filtering problems \cite{VP_2013a,VP_2013b}, first-order necessary conditions of optimality for the control problem (\ref{opt}) can be obtained by equating to zero the Frechet derivatives of $\Ups$  with respect to the controller matrices. A frequency-domain representation of these derivatives is provided by the following theorem. For its formulation, we will  use the auxiliary real matrices
\begin{equation}
\label{KK}
    K_1
    :=
  {\begin{bmatrix}
    0 & E\\
    I_n & 0\\
    0 & K
  \end{bmatrix}},
  \qquad
  K_2
  :=
  {\begin{bmatrix}
    0 & I_n & 0\\
    C & 0 & D
  \end{bmatrix}},
\end{equation}
specified by the state-space matrices of the quantum plant (\ref{dX}), the observation process (\ref{dZ}) and the weighting matrix $K$ in (\ref{cV}). Also,
\begin{align}
\label{phi}
    \phi
    := &
    \sinc(\theta \Psi) \Delta^{-1}, \\
\nonumber
    \psi
    := &
          \left(
          \sin
          {\begin{bmatrix}
            \theta \Psi & 0\\
            \Delta^{-1} \Phi\Psi^{-1} & \theta \Psi
          \end{bmatrix}}
            -
          \cos
          {\begin{bmatrix}
            \theta \Psi & 0\\
            \Delta^{-1} & \theta \Psi
          \end{bmatrix}}
          \right)_{21}\\
\label{psi}
          & -
          \sinc(\theta \Psi) \Delta^{-1}\Phi \Psi^{-1}
\end{align}
 are $\mC^{\nu\x \nu}$-valued functions of $\lambda \in \mR$ (the frequency argument is omitted for brevity) associated with (\ref{D})--(\ref{Psi0}), where $(\cdot)_{jk}$ is the $(j,k)$th block of an appropriately partitioned matrix. Furthermore, we will use the matrix
 \begin{equation}
 \label{chi}
   \chi
   :=
   \tfrac{1}{4\pi}
   \Re
   \int_{\mR}
   \fP
   \Big(
    {\small\begin{bmatrix}
      G \cB\\
      I_m
    \end{bmatrix}}
    (F^* (\phi \!+\! \phi^*) \!+\! J F^* (\psi \!-\! \psi^*))
    {\begin{bmatrix}
      \cC G & I_\nu
    \end{bmatrix}}
    \Big)
   \rd \lambda,
 \end{equation}
which is defined in terms of the matrix $J$ from (\ref{J}), the transfer functions (\ref{F0}), (\ref{G}) and the functions (\ref{phi}), (\ref{psi}). Here, $\fP$ is the orthogonal projection onto the subspace of matrices whose bottom right $(m\x\nu)$-block vanishes.
Since (\ref{psi}) involves $\Psi^{-1}$, it is assumed that (\ref{Psi0}) satisfies
\begin{equation}
\label{Psidet}
  \det \Psi(\lambda)\ne 0,
  \qquad
  \lambda \in \mR.
\end{equation}
Note that those matrix triples $(a,b,c)$, which specify stabilizing controllers (\ref{dxi}) satisfying (\ref{spec1}), (\ref{Psidet}), form an open subset of $\mR^{n\x n}\x \mR^{n\x r}\x\mR^{d \x n}$. Such controllers will be referred to as \emph{admissible} controllers.

\begin{thm}
\label{th:ders}
Suppose (\ref{dxi}) describes an admissible controller, and the closed-loop system (\ref{dcX})  is driven by vacuum fields. 
Then the partial Frechet derivatives of the QEF growth rate (\ref{Ups}) for the system in the invariant Gaussian state with respect to the controller matrices can be computed as
\begin{align}
\label{dUpsda}
    \d_a \Ups & = \theta(K_1^\rT \chi^\rT K_2^\rT)_{11},\\
\label{dUpsdb}
    \d_b \Ups & = \theta (K_1^\rT \chi^\rT K_2^\rT)_{12},\\
\label{dUpsdc}
    \d_c \Ups & = \theta (K_1^\rT \chi^\rT K_2^\rT)_{21}
\end{align}
in terms of (\ref{KK})--(\ref{chi}). \hfill$\square$
\end{thm}
\begin{pf}
In view of the relation $\delta \ln\det \gamma = \Tr(\gamma^{-1}\delta \gamma)$, it follows from (\ref{Ups}) that the first variation of the QEF growth rate with respect to the admissible controller matrices can be represented as
\begin{equation}
\label{dUps}
  \delta \Ups
  =
  -\tfrac{1}{4\pi}
  \int_{\mR}
  \Tr(\Delta(\lambda)^{-1}\delta \Delta(\lambda))
  \rd \lambda.
\end{equation}
For a fixed but otherwise arbitrary frequency $\lambda\in \mR$,   application of Lemma~\ref{lem:cos'sin'} from Appendix~\ref{sec:matfun} to the matrix $\Delta(\lambda)$ in (\ref{D}) leads to
\begin{align}
\nonumber
    \tfrac{1}{\theta}\delta  \Delta
    = &
- (\delta \Phi)\sinc(\theta \Psi)
     + \Phi \Psi^{-1} (\delta \Psi)\sinc(\theta \Psi)\\
\nonumber
    & +
    \tfrac{1}{\theta}
    (
    \delta\cos(\theta \Psi) - \Phi\Psi^{-1} \delta\sin(\theta \Psi))\\
\nonumber
    =& - (\delta \Phi)\sinc(\theta \Psi)
     + \Phi \Psi^{-1} (\delta \Psi)\sinc(\theta \Psi)\\
\label{dD}
    & +
    \left(
    \cos
    {\begin{bmatrix}
      \theta \Psi & 0\\
      \delta \Psi & \theta \Psi
    \end{bmatrix}}
    \right)_{21}
     - \Phi \Psi^{-1}
    \left(
    \sin
    {\begin{bmatrix}
      \theta \Psi & 0\\
      \delta \Psi & \theta \Psi
    \end{bmatrix}}
    \right)_{21}.
\end{align}
By combining (\ref{dD}) with the cyclic property of the matrix trace and 
Lemma~\ref{lem:fab}, it follows that
\begin{align}
\nonumber
    -\tfrac{1}{\theta}\Tr &(\Delta^{-1} \delta \Delta)\\
\nonumber
    =&  \Tr (\sinc(\theta \Psi)\Delta^{-1}\delta \Phi)\\
\nonumber
    & -
    \Tr(\sinc(\theta \Psi) \Delta^{-1}\Phi \Psi^{-1} \delta \Psi)\\
\nonumber
    &+
    \Tr
    \left(
          \left(
          \sin
          {\begin{bmatrix}
            \theta \Psi & 0\\
            \Delta^{-1} \Phi\Psi^{-1} & \theta \Psi
          \end{bmatrix}}
            -
          \cos
          {\begin{bmatrix}
            \theta \Psi & 0\\
            \Delta^{-1} & \theta \Psi
          \end{bmatrix}}
          \right)_{21}
          \delta \Psi
    \right)\\
\label{dlndet}
    =&
    \Tr(\phi \delta \Phi + \psi \delta\Psi),
\end{align}
where use is also made of (\ref{phi}), (\ref{psi}). Here, the first variations of the matrices $\Phi$, $\Psi$ in (\ref{Phi0}), (\ref{Psi0})  are given by
\begin{equation}
\label{dPhi_dPsi}
    \delta \Phi
    =
    (\delta F) F^* + F\delta F^*,
    \qquad
    \delta \Psi
    =
    (\delta F) J F^* + FJ\delta F^*.
\end{equation}
The first variation of the transfer matrix $F$ in (\ref{F0}) can be represented as
\begin{align}
\nonumber
  \delta F
  & =
  \cC G (\delta \cA) G \cB + \cC G \delta \cB + (\delta \cC) G \cB\\
\label{dF}
    & =
    {\begin{bmatrix}
      \cC G & I_\nu
    \end{bmatrix}}
  {\begin{bmatrix}
    \delta\cA & \delta\cB\\
    \delta\cC & 0
  \end{bmatrix}}
    {\begin{bmatrix}
      G \cB \\
      I_m
    \end{bmatrix}},
\end{align}
where use is made of the relation $\delta G = G(\delta \cA)G$,  which is obtained from (\ref{G})  and the identity $\delta (\gamma^{-1}) = -\gamma^{-1} (\delta \gamma) \gamma^{-1}$.
With the closed-loop system  matrices $\cA$, $\cB$, $\cC$ in (\ref{cAB}), (\ref{cC}) being affine functions of the controller matrices $a$, $b$, $c$ from (\ref{dxi}), their variations are related by
\begin{equation}
\label{dABC}
  {\begin{bmatrix}
    \delta\cA & \delta\cB\\
    \delta\cC & 0
  \end{bmatrix}}
  =
  K_1
  {\begin{bmatrix}
    \delta a & \delta b\\
    \delta c & 0
  \end{bmatrix}}
  K_2,
\end{equation}
where $K_1$, $K_2$ are given by (\ref{KK}), cf. \cite[Eqs. (22), (23)]{VP_2013a}. By substituting (\ref{dlndet})--(\ref{dABC}) into (\ref{dUps}), it follows that
\begin{align}
\nonumber
  \tfrac{1}{\theta}\delta &\Ups
  =
  \tfrac{1}{4\pi}
  \int_{\mR}
  \Tr(\phi \delta \Phi + \psi \delta\Psi)
  \rd \lambda\\
\nonumber
     =&
    \tfrac{1}{4\pi}
    \int_{\mR}
    \Tr(\phi ((\delta F) F^* \!+\! F\delta F^*) \!+\! \psi ((\delta F) J F^* \!+\! FJ\delta F^*))
    \rd \lambda\\
\nonumber
    =&
    \tfrac{1}{4\pi}
    \int_{\mR}
    \Tr
    (
        F^* \phi \delta F + \overline{F^* \phi^* \delta F}
        +
        J F^* \psi \delta F - J\overline{F^* \psi^* \delta F}
    )
    \rd \lambda\\
\nonumber
    =&
    \tfrac{1}{4\pi}
    \Re
    \int_{\mR}
    \Tr
    (
        (F^* (\phi+\phi^*) + J F^* (\psi-\psi^*))
        \delta F
    )
    \rd \lambda\\
\nonumber
    =&
    \tfrac{1}{4\pi}
    \Re
    \int_{\mR}
    \Tr
    \Big(
        (F^* (\phi+\phi^*) + J F^* (\psi-\psi^*))\\
\nonumber
& \x
            {\begin{bmatrix}
      \cC G & I_\nu
    \end{bmatrix}}
  {\begin{bmatrix}
    \delta \cA & \delta \cB\\
    \delta \cC & 0
  \end{bmatrix}}
    {\begin{bmatrix}
      G \cB \\
      I_m
    \end{bmatrix}}
    \Big)
    \rd \lambda\\
\nonumber
    =&
    \tfrac{1}{4\pi}\!
    \Re
    \Tr\!
    \Big(\!
    \int_{\mR}\!\!\!
  \fP\!
  \Big(\!
    {\begin{bmatrix}
      G \cB \\
      I_m
    \end{bmatrix}}\!
        (F^*\! (\phi\!+\!\phi^*) \!+\! J F^*\! (\psi\!-\!\psi^*))
        {\begin{bmatrix}
      \cC G & I_\nu
    \end{bmatrix}}\!
    \Big)\!
    \rd \lambda\\
\nonumber
& \x
  {\begin{bmatrix}
    \delta \cA & \delta \cB\\
    \delta \cC & 0
  \end{bmatrix}}
    \Big)\\
\label{dUps1}
    = &
    \Tr
    \Big(
        \chi
      {\begin{bmatrix}
        \delta \cA & \delta \cB\\
        \delta \cC & 0
      \end{bmatrix}}
    \Big)
    =
    \Tr
    \Big(
        K_2
        \chi
      K_1
      {\begin{bmatrix}
        \delta a & \delta b\\
        \delta c & 0
      \end{bmatrix}}
    \Big),
\end{align}
with $\chi$ given by (\ref{chi}).
Here, we have also used the cyclic property of the trace (and its invariance under the matrix transpose), the antisymmetry of the real matrix $J$ in (\ref{J}), the invariance of the real part under the complex conjugation, and the fact that both $\Ups$, the controller matrices $a$, $b$, $c$ and the matrices $K_1$, $K_2$ in (\ref{KK}) are real. Since the last trace in (\ref{dUps1}) is the Frobenius inner product \cite{HJ_2007} $\bra \cdot, \cdot\ket$ of the real matrices $(        K_2
        \chi
      K_1
)^\rT = K_1^\rT \chi^\rT K_2^\rT$ and $      {\small\begin{bmatrix}
        \delta a & \delta b\\
        \delta c & 0
      \end{bmatrix}}
$, it can be represented by using their block structure as
\begin{align}
\nonumber
    \tfrac{1}{\theta}\delta\Ups
    =&
    \bra (K_1^\rT \chi^\rT K_2^\rT)_{11}, \delta a \ket \\
\nonumber
    & +
    \bra (K_1^\rT \chi^\rT K_2^\rT)_{12}, \delta b \ket \\
\label{dUps2}
    & +
    \bra (K_1^\rT \chi^\rT K_2^\rT)_{21}, \delta c \ket.
\end{align}
Now, the controller matrices $a$, $b$, $c$ are independent variables, and hence, (\ref{dUps2}) leads to the partial Frechet derivatives (\ref{dUpsda})--(\ref{dUpsdc}).
\hfill$\blacksquare$
\end{pf}

The proof of Theorem~\ref{th:ders} shows that the matrix $\chi$ in (\ref{chi}) encodes the partial Frechet derivatives of the QEF growth rate $\Ups$ with respect to the closed-loop system matrices $\cA$, $\cB$, $\cC$ (regarded as independent variables):
\begin{equation}
\label{dUpsdABC}
    \theta \chi^\rT
    =
    {\begin{bmatrix}
      \d_{\cA} \Ups & \d_{\cB} \Ups\\
        \d_{\cC} \Ups & 0
    \end{bmatrix}}.
\end{equation}
Also, the factors $K_1^\rT$, $K_2^\rT$  in (\ref{dUpsda})--(\ref{dUpsdc}) originate from the fact that the adjoint of the linear operator $\[[[K_1, K_2\]]]: \mR^{(n+d)\x (n+r)}\ni \gamma\mapsto K_1 \gamma K_2 \in \mR^{(2n+\nu)\x(2n+m)}$ in (\ref{dABC})  is given by $\[[[K_1, K_2\]]]^\dagger = \[[[K_1^\rT, K_2^\rT\]]]$, with the Hilbert spaces of real matrices endowed with the Frobenius inner product.

As $\theta \to 0+$, the  functions (\ref{D}),  (\ref{phi}), (\ref{psi}) reduce to $\Delta = I_\nu$, $\phi = I_\nu$, $\psi=0$, and (\ref{chi}) yields the matrix
\begin{equation}
 \label{chi0}
   \chi_0
   :=
   \tfrac{1}{2\pi}
   \Re
   \int_{\mR}
   \fP
   \Big(
    {\begin{bmatrix}
      G(i\lambda) \cB\\
      I_m
    \end{bmatrix}}
    F(i\lambda)^*
    {\begin{bmatrix}
      \cC G(i\lambda) & I_\nu
    \end{bmatrix}}
    \Big)
   \rd \lambda,
 \end{equation}
which is related to the partial  Frechet derivatives of the LQG cost $\Ups_0 := 
\frac{1}{2} \bE (\cV(0)^\rT \cV(0)) = \lim_{\theta\to 0+} \big(\frac{1}{\theta} \Ups\big)$ for the closed-loop system in the invariant Gaussian state. In this limiting  case, in accordance with (\ref{dUpsdABC}),
\begin{equation}
\label{PQH}
    \chi_0^\rT
    =
    {\begin{bmatrix}
      \d_{\cA} \Ups_0 & \d_{\cB} \Ups_0\\
        \d_{\cC} \Ups_0 & 0
    \end{bmatrix}}
    =
    {\begin{bmatrix}
      \sH & \sQ\cB\\
      \cC \sP & 0
    \end{bmatrix}},
\end{equation}
where $\sH:= \sQ\sP$ is the Hankelian of the system \cite[Lemma~2]{VP_2013a} associated with its controllability and observability Gramians \cite{KS_1972} $\sP$, $\sQ$ satisfying the ALEs $\cA \sP + \sP \cA^\rT + \cB\cB^\rT = 0$ and $\cA^\rT \sQ + \sQ \cA + \cC^\rT\cC = 0$. The right-hand side of (\ref{PQH}) can also be obtained by evaluating (\ref{chi0}) through the Plancherel theorem applied to (\ref{F0}), (\ref{G}).

We now return to the case $\theta>0$.  In combination with (\ref{cAB}), (\ref{cC}), (\ref{KK})--(\ref{chi}), the equations
\begin{equation}
\label{optcond}
    (K_1^\rT \chi^\rT K_2^\rT)_{11}  =0,
    \qquad
    (K_1^\rT \chi^\rT K_2^\rT)_{12} = 0,
    \qquad
    (K_1^\rT \chi^\rT K_2^\rT)_{21} = 0
\end{equation}
provide first-order necessary conditions of optimality for the quantum risk-sensitive control problem (\ref{opt}) in the class of admissible measurement-based controllers (\ref{dxi}) for the quantum plant (\ref{dX}). 
Similarly to  \cite{SVP_2017},  these equations can be solved numerically by using the Frechet derivatives (\ref{dUpsda})--(\ref{dUpsdc}) for a gradient descent algorithm in the space of matrix triples $(a,b,c)$. A computationally challenging part of this approach is the evaluation of the matrix $\chi$ in (\ref{chi}) at every iteration. Since the state-space matrices of the measurement-based controller being considered are free from the PR constraints,  the gradient descent can be initialised with the standard classical LQG controller. Note that the state-space matrices of the LQG controller satisfy the optimality conditions (\ref{optcond}), where $\chi$ is replaced with the matrix $\chi_0$ from (\ref{PQH}).

\section{Conclusion}
\label{sec:conc}

We have considered a risk-sensitive optimal control problem for a quantum plant, where a stabilizing  measurement-based controller is sought to minimize the infinite-horizon growth rate of a quadratic-exponential cost for the closed-loop system in the invariant Gaussian state. In this setting, the controller influences the performance criterion not only through the statistical properties of the system variables but also through their two-point commutation structure.   We have obtained a frequency-domain representation for the Frechet derivatives of the cost functional with respect to the controller matrices, established first-order necessary conditions of optimality and outlined a gradient descent algorithm with the standard LQG controller as an initial approximation.

\appendix

\section{Differential identities for functions of matrices}\label{sec:matfun}
\renewcommand{\theequation}{A\arabic{equation}}
\setcounter{equation}{0}

The computation of Frechet derivatives in Section~\ref{sec:opt} employs the following lemmas concerning entire (and more specifically, trigonometric) functions of matrices.

\begin{lem}
\label{lem:fab}
Let $f: \mC\to \mC$ be an entire function with a globally convergent power series expansion $f(z) = \sum_{k=0}^{+\infty} f_k z^k$, where $f_k$ are complex coefficients. Then
\begin{equation}
\label{fab}
  \Tr (\alpha \delta f(\beta)) = \Tr (f'(\beta)[\alpha] \delta \beta)
\end{equation}
for any matrices $\alpha, \beta \in \mC^{\nu\x \nu}$,
where $\delta f(\beta)$ is the first variation of $f(\beta)$, and $f'(\beta)[\gamma]:= \lim_{\eps \to 0}\big(\frac{1}{\eps}(f(\beta + \eps \gamma)-f(\beta))\big)$ is the Gateaux derivative of $f$ at $\beta$ in the direction  $\gamma\in \mC^{\nu\x \nu}$. \hfill$\square$
\end{lem}
\begin{pf}
The first variation of $f(\beta)$ with respect to $\beta$ is represented as
\begin{equation}
\label{df}
    \delta f(\beta) = f'(\beta)[\delta \beta],
\end{equation}
where the Gateaux derivative is related to the
power series expansion of $f$ by
\begin{equation}
\label{f'}
    f'(\beta)[\gamma] = \sum_{k=0}^{+\infty} f_k \sum_{j=0}^{k-1} \beta^j \gamma \beta^{k-1-j}.
\end{equation}
In the case when $[\beta, \gamma]=0$, the representation (\ref{f'}) reduces to $f'(\beta)[\gamma] = f'(\beta)\gamma = \gamma f'(\beta)$, whereas $\gamma$ does not necessarily commute with $\beta$ in general.
By combining (\ref{df}) with (\ref{f'}), it follows that
\begin{align}
\nonumber
  \Tr
  (
    \alpha \delta f(\beta))
  & = \Tr (\alpha f'(\beta)[\delta \beta])
  )\\
\nonumber
  & =
  \Tr
  \Big(
    \alpha \sum_{k=0}^{+\infty} f_k \sum_{j=0}^{k-1} \beta^j (\delta \beta)\beta^{k-1-j}
  \Big)\\
\nonumber
  & =
  \Tr
  \Big(
    \sum_{k=0}^{+\infty} f_k \sum_{j=0}^{k-1} \beta^{k-1-j} \alpha \beta^j \delta \beta
    \Big)\\
\label{fab1}
    & = \Tr (f'(\beta)[\alpha] \delta \beta),
\end{align}
which establishes (\ref{fab}). The third equality in (\ref{fab1}) is obtained by using the cyclic property of the matrix trace.\hfill$\blacksquare$
\end{pf}

\begin{lem}
\label{lem:cos'sin'}
The Gateaux derivatives of the {\tt cos} and {\tt sin} functions, evaluated at a matrix $\beta \in \mC^{\nu \x\nu}$, can be computed as
\begin{align}
\label{cos'}
    (\cos\beta)'[\gamma]
    & =
    \left(
    \cos
    {\begin{bmatrix}
      \beta & 0\\
      \gamma & \beta
    \end{bmatrix}}
    \right)_{21},\\
\label{sin'}
    (\sin\beta)'[\gamma]
    & =
    \left(
    \sin
    {\begin{bmatrix}
      \beta & 0\\
      \gamma & \beta
    \end{bmatrix}}
    \right)_{21}
\end{align}
for any $\gamma \in \mC^{\nu\x \nu}$, where $(\cdot)_{21}$ is the bottom left $(\nu\x \nu)$-block  of a $(2\nu\x 2\nu)$-matrix. \hfill$\square$
\end{lem}
\begin{pf}
The Gateaux derivative of the matrix exponential is given by \cite{H_2008}
$$
    (\re^\beta)'[\gamma]
    =
    {\begin{bmatrix}
      0 & I_\nu
    \end{bmatrix}}
    \exp
    \left(
    {\begin{bmatrix}
      \beta & 0\\
      \gamma & \beta
    \end{bmatrix}}
    \right)
    {\begin{bmatrix}
      I_\nu\\
      0
    \end{bmatrix}    }
    =
    \left(
    \exp
    \left(
    {\begin{bmatrix}
      \beta & 0\\
      \gamma & \beta
    \end{bmatrix}}
    \right)
    \right)_{21}
$$
for any $\beta,\gamma \in \mC^{\nu\x \nu}$. Hence, the identities $\cos z = \frac{1}{2}(\re^{iz}+\re^{-iz})$ and $\sin z = \frac{1}{2i}(\re^{iz}-\re^{-iz})$ lead to
\begin{align*}
    (\cos\beta)'[\gamma]
    & =
    \tfrac{1}{2}
    \left(
    \exp
    \left(
    {\begin{bmatrix}
      i\beta & 0\\
      i\gamma & i\beta
    \end{bmatrix}}
    \right)
    +
    \exp
    \left(
    {\begin{bmatrix}
      -i\beta & 0\\
      -i\gamma & -i\beta
    \end{bmatrix}}
    \right)
    \right)_{21}\\
    & =
    \left(
    \cos
    {\begin{bmatrix}
      \beta & 0\\
      \gamma & \beta
    \end{bmatrix}}
    \right)_{21},\\
    (\sin\beta)'[\gamma]
    & =
    \tfrac{1}{2i}
    \left(
    \exp
    \left(
    {\begin{bmatrix}
      i\beta & 0\\
      i\gamma & i\beta
    \end{bmatrix}}
    \right)
    -
    \exp
    \left(
    {\begin{bmatrix}
      -i\beta & 0\\
      -i\gamma & -i\beta
    \end{bmatrix}}
    \right)
    \right)_{21}\\
    & =
    \left(
    \sin
    {\begin{bmatrix}
      \beta & 0\\
      \gamma & \beta
    \end{bmatrix}}
    \right)_{21},
\end{align*}
which establishes (\ref{cos'}), (\ref{sin'}). \hfill$\blacksquare$
\end{pf}
\end{document}